\documentstyle[preprint,aps]{revtex}

\begin{document}
\title{{\bf Approximate }$l${\bf -state} {\bf solutions of the }$D${\bf %
-dimensional Schr\"{o}dinger equation for Manning-Rosen potential }}
\author{Sameer M. Ikhdair\thanks{%
sameer@neu.edu.tr} and \ Ramazan Sever\thanks{%
sever@metu.edu.tr}}
\address{$^{\ast }$Department of Physics, Near East University, Nicosia, Cyprus,
Mersin 10, Turkey. \\
$^{\dagger }$Department of Physics, Middle East Technical University, 06531
Ankara, Turkey.}
\date{\today
}
\author{}
\maketitle

\begin{abstract}
The Schr\"{o}dinger equation in $D$-dimensions for the Manning-Rosen
potential with the centrifugal term is solved approximately to obtain bound
states eigensolutions (eigenvalues and eigenfunctions). The Nikiforov-Uvarov
(${\rm NU}$) method is used in the calculations. We present numerical
calculations of energy eigenvalues to two- and four-dimensional systems for
arbitrary quantum numbers $n$ and $l$ with three different values of the
potential parameter $\alpha .$ It is shown that because of the
interdimensional degeneracy of eigenvalues, we can also reproduce
eigenvalues of a upper/lower dimensional sytem from the well-known
eigenvalues of a lower/upper dimensional system by means of the
transformation $(n,l,D)\rightarrow (n,l\pm 1,D\mp 2).$. This solution
reduces to the Hulth\'{e}n potential case.


PACS NUMBER(S): 03.65.-w; 02.30.Gp; 03.65.Ge; 34.20.Cf
\end{abstract}

\bigskip

\section{Introduction}

\noindent One of the important tasks of quantum mechanics is to find exact
solutions of the wave equations (nonrelativistic and relativistic) for
certain potentials of physical interest since they contain all the necessary
information regarding the quantum system under consideration. It is well
known that the exact solutions of these wave equations are only possible in
a few simple cases such as the Coulomb, the harmonic oscillator,
pseudoharmonic and Mie-type potentials [1-8]. For an arbitrary $l$-state,
most quantum systems could be only treated by approximation methods. For the
rotating Morse potential some semiclassical and/or numerical solutions have
been obtained by using Pekeris approximation [9-13]. In recent years, many
authors have studied the nonrelativistic and relativistic wave equations
with certain potentials for the $s$- and $l$-cases. The exact and
approximate solutions of these models have been obtained analytically
[10-14].

Many exponential-type potentials have been solved like the Morse potential
[12,13,15], the Hulth\'{e}n potential [16-19], the P\"{o}schl-Teller [20],
the Woods-Saxon potential [21-23], the Kratzer-type potentials
[12,14,24-27], the Rosen-Morse-type potentials [28,29], the Manning-Rosen
potential [29-33] and other multiparameter exponential-type potentials
[34,35]. Various methods are used to obtain the exact solutions of the wave
equations for this type of exponential potentials (cf. [36] and the
references therein$)$.

Recently, the NU method [37] has shown its power in calculating the exact
energy levels of all bound states for some solvable quantum systems. In this
work, we attempt to apply this method to study another exponential-type
potential proposed by Manning and Rosen [29-33]. With an approximation to
centrifugal term, we solve the $D$-dimensional Schr\"{o}dinger equation to
its bound states energies and wavefunctions. This potential is defined as
[29-33]

\begin{equation}
V(r)=-V_{0}\frac{e^{-r/b}}{1-e^{-r/b}}+V_{1}\left( \frac{e^{-r/b}}{1-e^{-r/b}%
}\right) ^{2},\text{ }V_{0}=\frac{A}{\kappa b^{2}},\text{ }V_{1}=\frac{%
\alpha (\alpha -1)}{\kappa b^{2}},\text{ }\kappa =2\mu /\hbar ^{2},
\end{equation}
where $A$ and $\alpha $ are two-dimensionless parameters [27,28] but the
screening parameter $b$ has dimension of length which has a potential range $%
1/b.$ The potential (1) may be further put in the following simple form

\begin{equation}
V(r)=-\frac{Ce^{-r/b}+De^{-2r/b}}{\left( 1-e^{-r/b}\right) ^{2}},\text{ }C=A,%
\text{ }D=-A-\alpha \text{(}\alpha \text{-1),}
\end{equation}
which is usually used for the description of diatomic molecular vibrations
[38,39]. It is also used in several branches of physics for their bound
states and scattering properties. The potential in (1) remains invariant by
mapping $\alpha \rightarrow 1-\alpha $ and has a relative minimum value $%
V(r_{0})=-\frac{A^{2}}{4\kappa b^{2}\alpha (\alpha -1)}$ at $r_{0}=b\ln %
\left[ 1+\frac{2\alpha (\alpha -1)}{A}\right] $ for $\alpha >0$ to be
obtained from the first derivative $\left. \frac{dV}{dr}\right|
_{r=r_{0}}=0. $ The second derivative which determines the force constants
at $r=r_{0}$ is given by

\begin{equation}
\left. \frac{d^{2}V}{dr^{2}}\right| _{r=r_{0}}=\frac{A^{2}\left[ A+2\alpha
(\alpha -1)\right] ^{2}}{8b^{4}\alpha ^{3}(\alpha -1)^{3}}.
\end{equation}
The contents of this paper are as follows: In Section II we outline the
Nikiforov-Uvarov (NU) method. In Section III, we derive $l\neq 0$ bound
state eigensolutions of the $D$-dimensional Schr\"{o}dinger equation for the
Manning-Rosen potential by this method. In Section IV, we present our
numerical calculations in $2D$ and $4D$ systems for various quantum numbers $%
n$ and $l.$ Section V, is devoted to for two special cases, namely, $l=0$
and the Hulth\'{e}n potential. The concluding remarks are given in Section
VI.

\section{\noindent The Nikiforov-Uvarov method}

The NU method is based on solving the second-order linear differential
equation by reducing it to a generalized equation of hypergeometric type
[37]. In this method after employing an appropriate coordinate
transformation $z=z(r),$ the Schr\"{o}dinger equation can be written in the
following form:
\[
\psi _{n}^{\prime \prime }(z)+\frac{\widetilde{\tau }(z)}{\sigma (z)}\psi
_{n}^{\prime }(z)+\frac{\widetilde{\sigma }(z)}{\sigma ^{2}(z)}\psi
_{n}(z)=0,
\]
\begin{equation}
\psi _{n}(z)=\phi _{n}(z)y_{n}(z),
\end{equation}
where $\sigma (z)$ and $\widetilde{\sigma }(z)$ are the polynomials with at
most of second-degree, and $\widetilde{\tau }(s)$ is a first-degree
polynomial. The special orthogonal polynomials [37] reduce Eq. (4) to a
simple equation of the following hypergeometric type:

\begin{equation}
\sigma (z)y_{n}^{\prime \prime }(z)+\tau (z)y_{n}^{\prime }(z)+\lambda
y_{n}(z)=0,
\end{equation}
where

\[
\tau (z)=\widetilde{\tau }(z)+2\pi (z),\text{ }\tau ^{\prime }(z)<0,
\]
\begin{equation}
\sigma (z)=\pi (z)\frac{\phi (z)}{\phi ^{\prime }(z)},
\end{equation}
and $\lambda $ is a constant given in the form

\[
\lambda =\lambda _{n}=-n\tau ^{\prime }(z)-\frac{n\left( n-1\right) }{2}%
\sigma ^{\prime \prime }(z),\text{\ \ \ }n=0,1,2,...
\]
\begin{equation}
\lambda =k+\pi ^{\prime }(z).
\end{equation}
It is worthwhile to note that $\lambda $ or $\lambda _{n}$ are obtained from
a particular solution of the form $y(z)=y_{n}(z)$ which is a polynomial of
degree $n.$ Further, $\ y_{n}(z)$ is the hypergeometric-type function whose
polynomial solutions are given by Rodrigues relation

\begin{equation}
y_{n}(z)=\frac{B_{n}}{\rho (z)}\frac{d^{n}}{dz^{n}}\left[ \sigma ^{n}(z)\rho
(z)\right] ,
\end{equation}
where $B_{n}$ is the normalization constant and the weight function $\rho
(z) $ must satisfy the condition [37]

\begin{equation}
\frac{d}{dz}w(z)=\frac{\tau (z)}{\sigma (z)}w(z),\text{ }w(z)=\sigma (z)\rho
(z).
\end{equation}
In order to determine the weight function given in Eq. (9), we must obtain
the following polynomial:

\begin{equation}
\pi (z)=\frac{\sigma ^{\prime }(z)-\widetilde{\tau }(z)}{2}\pm \sqrt{\left(
\frac{\sigma ^{\prime }(z)-\widetilde{\tau }(z)}{2}\right) ^{2}-\widetilde{%
\sigma }(z)+k\sigma (z)}.
\end{equation}
In principle, the expression under the square root sign in Eq. (10) can be
arranged as the square of a polynomial. This is possible only if its
discriminant is zero. In this case, an equation for $k$ is obtained. After
solving this equation, the obtained values of $k$ are included in the ${\rm %
NU}$ method and here there is a relationship between $\lambda $ and $k$
given in Eq. (7).

\section{Bound-state solutions for arbitrary $l$-state}

In this section, we follow closely the approach of Ref. [36]. We begin by
considering the SE, in arbitrary dimension $D,$ as [40-42]

\[
\left\{ \nabla _{D}^{2}+\frac{2\mu }{\hbar ^{2}}\left[ E_{nl}-V(r)\right]
\right\} \psi _{l_{1}\cdots l_{D-2}}^{(l_{D-1}=l)}({\bf x})=0,
\]
\[
\nabla _{D}^{2}=\frac{\partial ^{2}}{\partial r^{2}}+\frac{(D-1)}{r}\frac{%
\partial }{\partial r}
\]
\[
+\frac{1}{r^{2}}\left[ \frac{1}{\sin ^{D-2}\theta _{D-1}}\frac{\partial }{%
\partial \theta _{D-1}}\left( \sin ^{D-2}\theta _{D-1}\frac{\partial }{%
\partial \theta _{D-1}}\right) -\frac{L_{D-2}^{2}}{\sin ^{2}\theta _{D-1}}%
\right] ,
\]
\begin{equation}
\psi _{l_{1}\cdots l_{D-2}}^{(l)}({\bf x})=R_{l}(r)Y_{l_{1}\cdots
l_{D-2}}^{(l)}(\widehat{{\bf x}}),\text{ }R_{l}(r)=r^{-(D-1)/2}g(r),\text{ }
\end{equation}
where the potential $V(r)$ is taken as the Manning-Rosen form in (1). In
addition, $\mu $ $\ $and $E_{nl}$ denote the reduced mass and energy of two
interacting particles, respectively. ${\bf x}$ is a $D$-dimensional position
vector with the hyperspherical Cartesian components $x_{1},x_{2},\cdots
,x_{D}$ given as follows [43-47]:

\[
x_{1}=r\cos \theta _{1}\sin \theta _{2}\cdots \sin \theta _{D-1},
\]
\[
x_{2}=r\sin \theta _{1}\sin \theta _{2}\cdots \sin \theta _{D-1},
\]
\[
x_{3}=r\cos \theta _{2}\sin \theta _{3}\cdots \sin \theta _{D-1},
\]
\[
\vdots
\]
\[
x_{j}=r\cos \theta _{j-1}\sin \theta _{j}\cdots \sin \theta _{D-1},\text{ }%
3\leq j\leq D-1,
\]
\[
\vdots
\]
\[
x_{D-1}=r\cos \theta _{D-2}\sin \theta _{D-1},
\]
\begin{equation}
x_{D}=r\cos \theta _{D-1},\text{ }\sum\limits_{j=1}^{D}x_{j}^{2}=r^{2},
\end{equation}
for $D=2,3,\cdots .$ We have $x_{1}=r\cos \varphi ,$ $x_{2}=r\sin \varphi $
for $D=2$ and $x_{1}=r\cos \varphi \sin \theta ,$ $x_{2}=r\sin \varphi \sin
\theta ,$ $x_{3}=r\cos \theta $ for $D=3.$ The Laplace operator $\nabla
_{D}^{2}$ is defined by [48]
\begin{equation}
\nabla _{D}^{2}=\sum\limits_{j=1}^{D}\frac{\partial ^{2}}{\partial x_{j}^{2}}%
.
\end{equation}
The volume element of the configuration space is given by
\begin{equation}
\prod\limits_{j=1}^{D}dx_{j}=r^{D-1}drd\Omega ,\text{ }d\Omega
=\prod\limits_{j=1}^{D-1}(\sin \theta _{j})^{j-1}d\theta _{j},
\end{equation}
where $r\in \lbrack 0,\infty ),$ $\theta _{1}\in \lbrack 0,2\pi ]$ and $%
\theta _{j}\in \lbrack 0,\pi ],$ $j\in \lbrack 2,D-1].$ Equation (13)
permits a solution via separation of variables, if one writes generalized
spherical harmonics $Y_{\ell _{1}\cdots \ell _{D-2}}^{(\ell )}(\widehat{{\bf %
x}})$ [49]:

\[
Y_{l_{1}\cdots l_{D-2}}^{(l)}(\widehat{{\bf x}})=Y(l_{1},l_{2},\cdots
,l_{D-2},l),\text{ }l=\left| m\right| \text{ for }D=2,
\]
\begin{equation}
Y_{l_{1}\cdots l_{D-2}}^{(l)}(\widehat{{\bf x}}=\theta _{1},\theta
_{2},\cdots ,\theta _{D-1})=\prod\limits_{j=1}^{D-1}H(\theta _{j}),
\end{equation}
as a simultaneous eigenfunction of $L_{j}^{2}:$%
\[
L_{1}^{2}Y_{l_{1}\cdots l_{D-2}}^{(l)}(\widehat{{\bf x}})=m^{2}Y_{l_{1}%
\cdots l_{D-2}}^{(l)}(\widehat{{\bf x}}),
\]
\[
L_{j}^{2}Y_{l_{1}\cdots l_{D-2}}^{l\ell )}(\widehat{{\bf x}}%
)=l_{j}(l_{j}+j-1)Y_{l_{1}\cdots l_{D-2}}^{(l)}(\widehat{{\bf x}}),\text{ }%
j\in \lbrack 1,D-1],
\]
\[
l=0,1,\cdots ,l_{k}=0,1,\cdots ,l_{k+1},\text{ }k\in \lbrack 2,D-2],
\]
\[
l_{1}=-l_{2},-l_{2}+1,\cdots ,l_{2}-1,l_{2},
\]
\begin{equation}
L_{D-1}^{2}Y_{l_{1}\cdots l_{D-2}}^{(l)}(\widehat{{\bf x}}%
)=l(l+D-2)Y_{l_{1}\cdots l_{D-2}}^{(l)}(\widehat{{\bf x}}).
\end{equation}
The unit vector along ${\bf x}$ is usually denoted by $\widehat{{\bf x}}=%
{\bf x}/r{\bf .}$ Additionally, the angular momentum operators $L_{j}^{2}$
are defined as [43-47,49]$:$%
\[
L_{1}^{2}=-\frac{\partial ^{2}}{\partial \theta _{1}^{2}},
\]
\[
L_{k}^{2}=\sum\limits_{a<b=2}^{k+1}L_{ab}^{2}=-\frac{1}{\sin ^{k-1}\theta
_{k}}\frac{\partial }{\partial \theta _{k}}\left( \sin ^{k-1}\theta _{k}%
\frac{\partial }{\partial \theta _{k}}\right) +\frac{L_{k-1}^{2}}{\sin
^{2}\theta _{k}},\text{ }2\leq k\leq D-1,
\]
\begin{equation}
L_{ab}=-i\left[ x_{a}\frac{\partial }{\partial x_{b}}-x_{b}\frac{\partial }{%
\partial x_{a}}\right] .
\end{equation}
The substitution of Eqs. (13) and (15)-(17) into Eq. (11) allows us to
obtain, via the method of separation of variables, the following equation:
\[
\left\{ \frac{\partial ^{2}}{\partial r^{2}}+\frac{(D-1)}{r}\frac{\partial }{%
\partial r}+\frac{1}{r^{2}}\frac{1}{\sin ^{D-2}\theta _{D-1}}\frac{\partial
}{\partial \theta _{D-1}}\left( \sin ^{D-2}\theta _{D-1}\frac{\partial }{%
\partial \theta _{D-1}}\right) -\frac{L_{D-2}^{2}}{\sin ^{2}\theta _{D-1}}%
\right.
\]
\[
+\left. \frac{2\mu }{\hbar ^{2}}E_{nl}-\frac{\alpha (\alpha -1)}{b^{2}}\frac{%
e^{-2r/b}}{\left( 1-e^{-r/b}\right) ^{2}}+\frac{A}{b^{2}}\frac{e^{-r/b}}{%
1-e^{-r/b}}-\frac{l(l+D-2)}{r^{2}}\right\}
\]
\[
\times r^{-(D-1)/2}g(r)Y_{l_{1}\cdots l_{D-2}}^{(l)}(\widehat{{\bf x}})=0,
\]
\[
-L_{D-1}^{2}=\frac{1}{\sin ^{D-2}\theta _{D-1}}\frac{\partial }{\partial
\theta _{D-1}}\left( \sin ^{D-2}\theta _{D-1}\frac{\partial }{\partial
\theta _{D-1}}\right) -\frac{L_{D-2}^{2}}{\sin ^{2}\theta _{D-1}},
\]

\[
-L_{D-2}^{2}=\frac{1}{\sin ^{D-3}\theta _{D-2}}\frac{\partial }{\partial
\theta _{D-2}}\left( \sin ^{D-3}\theta _{D-2}\frac{\partial }{\partial
\theta _{D-2}}\right) -\frac{L_{D-3}^{2}}{\sin ^{2}\theta _{D-2}},
\]
\[
\vdots
\]
\[
-L_{j}^{2}=\frac{1}{\sin ^{j-1}\theta _{j}}\frac{\partial }{\partial \theta
_{j}}\left( \sin ^{j-1}\theta _{j}\frac{\partial }{\partial \theta _{j}}%
\right) -\frac{L_{j-1}^{2}}{\sin ^{2}\theta _{j}},\text{ \ }j\in \lbrack
2,D-2],
\]
\[
\vdots
\]
\begin{equation}
-L_{1}^{2}=\frac{\partial ^{2}}{\partial \theta _{1}^{2}},
\end{equation}
where $L_{k}^{2},$ \ $k\in \lbrack 1,D-1]$ are the angular operators. Thus,
with the aid of Eqs. (15) and (16), the last wave equation can be easily
separated into the following radial and angular parts as [50,51]:\footnote{%
The Schr\"{o}dinger equation in the presence of this potential is separable
to $(D-1)$- angular equations for the angular parameters $(\theta _{1}=\phi
,\theta _{2},...,\theta _{D-1}=\theta )$ and one radial equation for the
radial parameter $r$ with calculated separation constants $\Lambda _{p}$
where $p\in \lbrack 1,D-1].$}

\[
\left\{ \frac{d^{2}}{dr^{2}}+\frac{2\mu }{\hbar ^{2}}E_{nl}-V_{eff}(r)\right%
\} g(r)=0,
\]
\begin{equation}
V_{eff}(r)=\frac{1}{b^{2}}\left[ \frac{\alpha (\alpha -1)e^{-2r/b}}{\left(
1-e^{-r/b}\right) ^{2}}-\frac{Ae^{-r/b}}{1-e^{-r/b}}\right] +\frac{%
(D+2l-2)^{2}-1}{4r^{2}},
\end{equation}
\[
\left[ \frac{1}{\sin ^{D-2}\theta _{D-1}}\frac{d}{d\theta _{D-1}}\left( \sin
^{D-2}\theta _{D-1}\frac{d}{d\theta _{D-1}}\right) +l(l+D-2)\right.
\]
\begin{equation}
-\left. \frac{\Lambda _{D-2}}{\sin ^{2}\theta _{D-1}}\right] H(\theta
_{D-1})=0,
\end{equation}
\[
\vdots
\]
\begin{equation}
\left[ \frac{1}{\sin ^{j-1}\theta _{j}}\frac{d}{d\theta _{j}}\left( \sin
^{j-1}\theta _{j}\frac{d}{d\theta _{j}}\right) +\Lambda _{j}-\frac{\Lambda
_{j-1}}{\sin ^{2}\theta _{j}}\right] H(\theta _{j})=0,\text{ \ }j\in \lbrack
2,D-2],
\end{equation}
\[
\vdots
\]
\begin{equation}
\left[ \frac{d^{2}}{d\theta _{1}^{2}}+\Lambda _{1}\right] H(\theta _{1})=0,%
\text{ }\Lambda _{1}=l_{1}^{2}=m^{2},
\end{equation}
where $\Lambda _{p}=l_{p}(l_{p}+p-1),$ $p\in \lbrack 1,D-1]$ are separation
constants and $g(r)$ is defined in Eq. (11). The solution in (22) is
periodic and must satisfy the periodic boundary condition $H(\theta
_{1}=\varphi )=H(\theta _{1}=\varphi +2\pi )$ from which we obtain [42]
\begin{equation}
H(\theta _{1})=\frac{1}{\sqrt{2\pi }}\exp (\pm il_{1}\theta _{1}),\text{ \ }%
l_{1}=0,1,2,\cdots .
\end{equation}
Further, Eqs. (20) and (21) representing the angular wave equation become
\[
\frac{d^{2}H(\theta _{D-1})}{d\theta _{D-1}^{2}}+(D-2)\frac{\cos \theta
_{D-1}}{\sin \theta _{D-1}}\frac{dH(\theta _{D-1})}{d\theta _{D-1}}
\]
\begin{equation}
+\left[ l(l+D-2)-\frac{\Lambda _{D-2}}{\sin ^{2}\theta _{D-1}}\right]
H(\theta _{D-1})=0,
\end{equation}
\begin{equation}
\frac{d^{2}H(\theta _{j})}{d\theta _{j}^{2}}+(j-1)\frac{\cos \theta _{j}}{%
\sin \theta _{j}}\frac{dH(\theta _{j})}{d\theta _{j}}+\left( \Lambda _{j}-%
\frac{\Lambda _{j-1}}{\sin ^{2}\theta _{j}}\right) H(\theta _{j})=0,
\end{equation}
with $j\in \lbrack 2,D-2],$ $D>3$ and $\Lambda _{p}$ which is well-known in
three-dimensional space [48]$.$\footnote{$\Lambda _{D-2}=m^{2}$ and $%
l_{D-2}=m$ for $D=3.$} Hence, Eqs. (24) and (25) are to be solved in the
following subsection.

\subsection{The solutions of the $D$-dimensional angular equations}

In order to apply NU method, we introduce a new variable $s=\cos \theta
_{j}. $ Hence, Eq. (25) is then rearranged in the form of the universal
associated-Legendre differential equation

\begin{equation}
\frac{d^{2}H(s)}{ds^{2}}-\frac{js}{1-s^{2}}\frac{dH(s)}{ds}+\frac{\Lambda
_{j}-\Lambda _{j-1}-\Lambda _{j}s^{2}}{(1-s^{2})^{2}}H(s)=0,
\end{equation}
where $j\in \lbrack 2,D-2],$ $D>3.$ By comparing Eqs. (26) and (4), the
corresponding polynomials are obtained
\begin{equation}
\widetilde{\tau }(s)=-js,\text{ \ \ \ }\sigma (s)=1-s^{2},\text{ \ \ }%
\widetilde{\sigma }(s)=-\Lambda _{j}s^{2}+\Lambda _{j}-\Lambda _{j-1}.
\end{equation}
Inserting the above expressions into Eq. (10) and taking $\sigma ^{\prime
}(s)=-2s$, one obtains the following function:

\begin{equation}
\pi (s)=\frac{(j-2)}{2}s\pm \sqrt{\left[ \left( \frac{j-2}{2}\right)
^{2}+\Lambda _{j}-k\right] s^{2}+k-\Lambda _{j}+\Lambda _{j-1}}.
\end{equation}
Following the method, the polynomial $\pi (s)$ is found to have the
following four possible values:
\begin{equation}
\pi (s)=\left\{
\begin{array}{cc}
\left( \frac{j-2}{2}+\widetilde{\Lambda }_{j-1}\right) s & \text{\ for }%
k_{1}=\Lambda _{j}-\Lambda _{j-1}, \\
\left( \frac{j-2}{2}-\widetilde{\Lambda }_{j-1}\right) s & \text{\ for }%
k_{1}=\Lambda _{j}-\Lambda _{j-1}, \\
\frac{(j-2)}{2}s+\widetilde{\Lambda }_{j-1} & \text{\ for }k_{2}=\Lambda
_{j}+\left( \frac{j-2}{2}\right) ^{2}, \\
\frac{(j-2)}{2}s-\widetilde{\Lambda }_{j-1} & \text{\ for }k_{2}=\Lambda
_{j}+\left( \frac{j-2}{2}\right) ^{2},
\end{array}
\right.
\end{equation}
where $\widetilde{\Lambda }_{p}=l_{p}+(p-1)/2,$ with $p=j-1,$ $j$ and $j\in
\lbrack 2,D-2],$ $D>3\ .$ Imposing the condition $\tau ^{\prime }(s)<0$ in
Eq. (6), one selects the following physically valid solutions:

\begin{equation}
k_{1}=\Lambda _{j}-\Lambda _{j-1}\text{\ \ and \ \ }\pi (s)=\left( \frac{j-2%
}{2}-\widetilde{\Lambda }_{j-1}\right) s,
\end{equation}
which yields from Eq. (6) that
\begin{equation}
\tau (s)=-2(1+\widetilde{\Lambda }_{j-1})s.
\end{equation}
Making use from Eq. (7), the following expressions for $\lambda $ are
obtained as follows:

\begin{equation}
\lambda =\lambda _{n_{j}}=2n_{j}(1+\widetilde{\Lambda }%
_{j-1})+n_{j}(n_{j}-1),
\end{equation}
\begin{equation}
\lambda =\Lambda _{j}-\Lambda _{j-1}-\widetilde{\Lambda }_{j-1}+\frac{j-2}{2}%
.
\end{equation}
Upon comparing Eqs. (32) and (33), we obtain

\begin{equation}
n_{j}=\widetilde{\Lambda }_{j}-\widetilde{\Lambda }_{j-1}-\frac{1}{2}.
\end{equation}
In addition, using Eqs. (6) and (8)-(9), we obtain the following useful
parts of the wavefunctions:

\begin{equation}
\phi (s)=\left( 1-s^{2}\right) ^{l_{j-1}/2},\text{ }\rho (s)=\left(
1-s^{2}\right) ^{\widetilde{\Lambda }_{j-1}},
\end{equation}
where $j\in \lbrack 2,D-2],$ $D>3.$ Besides, substituting the weight
function $\rho (s)$ given in (35) into Eq. (8), we obtain
\begin{equation}
y_{n_{j}}(s)=A_{n_{j}}\left( 1-s^{2}\right) ^{-\widetilde{\Lambda }_{j-1}}%
\frac{d^{n_{j}}}{ds^{n_{j}}}\left( 1-s^{2}\right) ^{n_{j}+\widetilde{\Lambda
}_{j-1}},
\end{equation}
where $A_{n_{j}}$ is the normaliation factor. Finally the angular
wavefunction is
\begin{equation}
H_{n_{j}}(\theta _{j})=N_{n_{j}}\left( \sin \theta _{j}\right)
^{l_{j-1}}P_{n_{j}}^{(\widetilde{\Lambda }_{j-1},\widetilde{\Lambda }%
_{j-1})}(\cos \theta _{j}),\text{ }j\in \lbrack 2,D-2],\text{ }D>3
\end{equation}
with $n_{j}$ given in Eq. (34) becomes
\begin{equation}
n_{j}=l_{j}-l_{j-1},\text{ }j\in \lbrack 2,D-2],D>3.
\end{equation}
Likewise, using $s=\cos \theta _{D-1},$ we can rewrite Eq. (24) in the
associated Legendre form

\begin{equation}
\frac{d^{2}H(s)}{ds^{2}}-\frac{(D-1)s}{1-s^{2}}\frac{dH(s)}{ds}+\frac{\nu
(1-s^{2})-\Lambda _{D-2}}{(1-s^{2})^{2}}H(s)=0,
\end{equation}
\begin{equation}
\nu =l(l+D-2).
\end{equation}
It's worth to note that, Eq. (39) has been recently solved in $3D$ by the NU
method in [40,41,51]. However, our aim is to solve it in $D$-dimensions.
Hence, comparing Eqs. (39) and (4), the corresponding polynomials are
obtained
\begin{equation}
\widetilde{\tau }(s)=-(D-1)s,\text{ \ \ \ }\sigma (s)=1-s^{2},\text{ \ \ }%
\widetilde{\sigma }(s)=-\nu s^{2}+\nu -\Lambda _{D-2}.
\end{equation}
Inserting the above expressions into Eq. (10) and taking $\sigma ^{\prime
}(s)=-2s$, one obtains:

\begin{equation}
\pi (s)=\frac{(D-3)}{2}s\pm \sqrt{\left[ \left( \frac{D-3}{2}\right)
^{2}+\nu -k\right] s^{2}+k-\nu +\Lambda _{D-2}}.
\end{equation}
Following the method, the polynomial $\pi (s)$ is found to have the
following four possible values:
\begin{equation}
\pi (s)=\left\{
\begin{array}{cc}
\left( \frac{D-3}{2}+\widetilde{\Lambda }_{D-2}\right) s & \text{\ for }%
k_{1}=\nu -\Lambda _{D-2}, \\
\left( \frac{D-3}{2}-\widetilde{\Lambda }_{D-2}\right) s & \text{\ for }%
k_{1}=\nu -\Lambda _{D-2}, \\
\frac{(D-3)}{2}s+\widetilde{\Lambda }_{D-2} & \text{\ for }k_{2}=\nu +\left(
\frac{D-3}{2}\right) ^{2}, \\
\frac{(D-3)}{2}s-\widetilde{\Lambda }_{D-2} & \text{\ for }k_{2}=\nu +\left(
\frac{D-3}{2}\right) ^{2},
\end{array}
\right.
\end{equation}
where $\widetilde{\Lambda }_{D-2}=l_{D-2}+\frac{D-3}{2}.$ Imposing the
condition $\tau ^{\prime }(s)<0$ in Eq. (6), one selects the following
physically valid solutions:

\begin{equation}
k_{1}=\nu -\Lambda _{D-2}\text{ \ \ and \ \ }\pi (s)=-l_{D-2}s,\text{ }\nu
=l(l+D-2),
\end{equation}
giving
\begin{equation}
\tau (s)=-2(1+\widetilde{\Lambda }_{D-2})s.
\end{equation}
Making use from Eq. (7), we obtain
\begin{equation}
\lambda =\lambda _{n_{D-1}}=2n_{D-1}(1+\widetilde{\Lambda }%
_{D-2})+n_{D-1}(n_{D-1}-1),
\end{equation}
\begin{equation}
\lambda =l(l+D-2)-l_{D-2}(l_{D-2}+D-2).
\end{equation}
We compare Eqs. (46) and (47), the angular momentum\ $l$ values are obtained
from

\begin{equation}
l=n_{D-1}+l_{D-2},
\end{equation}
which can be easily reduced to the well-known solution
\begin{equation}
l=n+m,
\end{equation}
in $3D$ [51]. Using Eqs (6) and (8)-(9), we obtain the following useful
parts of the wavefunctions:

\begin{equation}
\phi (s)=\left( 1-s^{2}\right) ^{l_{D-2}/2},\text{ }\rho (s)=\left(
1-s^{2}\right) ^{\widetilde{\Lambda }_{D-2}}.
\end{equation}
Besides, the Rodrigues relation (8) gives
\begin{equation}
y_{n_{D-1}}(s)=B_{n_{D-1}}\left( 1-s^{2}\right) ^{-\widetilde{\Lambda }%
_{D-2}}\frac{d^{n_{D-1}}}{ds^{n_{D-1}}}\left( 1-s^{2}\right) ^{n_{D-1}+%
\widetilde{\Lambda }_{D-2}},
\end{equation}
where $B_{n_{D-1}}$ is the normaliation factor. Finally the angular
wavefunctions are
\begin{equation}
H_{n_{D-1}}(\theta _{D-1})=N_{n_{D-1}}\left( \sin \theta _{D-1}\right)
^{l_{D-2}}P_{n_{D-1}}^{(\widetilde{\Lambda }_{D-2},\widetilde{\Lambda }%
_{D-2})}(\cos \theta _{D-1}),
\end{equation}
where $n_{D-1}$ is given by Eq. (48).

\subsection{The solutions of the $D$-dimensional radial equation}

Since the radial part of \ the $D$-dimensional Schr\"{o}dinger equation with
above Manning-Rosen effective potential has no analytical solution for $%
l\neq 0$ states$,$ an approximation to the centrifugal term has to be made.
The good approximation for $1/r^{2}$ in the centrifugal barrier is taken as
[18,33]\footnote{%
The series approximation to the expression $\frac{1}{b^{2}}\frac{e^{-2r/b}}{%
\left( 1-e^{-r/b}\right) ^{2}}\approx \frac{1}{r^{2}}-\frac{1}{br},$ it
includes a Coulomb term.}

\begin{equation}
\frac{1}{r^{2}}\approx \frac{1}{b^{2}}\frac{e^{-r/b}}{\left(
1-e^{-r/b}\right) ^{2}},
\end{equation}
in a short potential range. To solve it by the present method, we need to
recast Eq. (19) with the aid of Eq. (53), into the form of Eq. (4) changing
the variables $r\rightarrow z$ through the mapping function $r=f(z)$ and
energy transformation given by

\begin{equation}
z=e^{-r/b},\text{ }\varepsilon =\sqrt{-\frac{2\mu b^{2}E_{nl}}{\hbar ^{2}}},%
\text{ }E_{nl}<0,
\end{equation}
to obtain the following hypergeometric equation:

\[
\frac{d^{2}g((z)}{dz^{2}}+\frac{(1-z)}{z(1-z)}\frac{dg(z)}{dz}
\]

\begin{equation}
+\frac{1}{\left[ z(1-z)\right] ^{2}}\left\{ -\varepsilon ^{2}+\frac{1}{4}%
\left[ 4A+8\varepsilon ^{2}-(D+2l-2)^{2}+1\right] z-\left[ A+\varepsilon
^{2}+\alpha (\alpha -1)\right] z^{2}\right\} g(z)=0.
\end{equation}
We notice that for bound state (real) solutions, the last equation requires
that

\begin{equation}
z=\left\{
\begin{array}{ccc}
0, & \text{when} & r\rightarrow \infty , \\
1, & \text{when} & r\rightarrow 0,
\end{array}
\right.
\end{equation}
and thus the finite radial wavefunctions $R_{nl}(z)\rightarrow 0.$ To apply
the NU method, it is necessary to compare Eq. (15) with Eq. (4).
Subsequently, the following value for the parameters in Eq. (4) are obtained
as
\[
\widetilde{\tau }(z)=1-z,\text{\ }\sigma (z)=z-z^{2}
\]
\begin{equation}
\widetilde{\sigma }(z)=-\left[ A+\varepsilon ^{2}+\alpha (\alpha -1)\right]
z^{2}+\frac{1}{4}\left[ 4A+8\varepsilon ^{2}-(D+2l-2)^{2}+\right]
z-\varepsilon ^{2}.
\end{equation}
If one inserts these values of parameters into Eq. (10), with $\sigma
^{\prime }(z)=1-2z,$ the following linear function is achieved
\[
\pi (z)=-\frac{z}{2}
\]
\begin{equation}
\pm \frac{1}{2}\sqrt{\left\{ 1+4\left[ A+\varepsilon ^{2}+\alpha (\alpha -1)%
\right] -k\right\} z^{2}+\left\{ 4k-\left[ 4A+8\varepsilon
^{2}-(D+2l-2)^{2}+1\right] \right\} z+4\varepsilon ^{2}}.
\end{equation}
According to this method the expression in the square root has to be set
equal to zero, that is, $\Delta =\left\{ 1+4\left[ A+\varepsilon ^{2}+\alpha
(\alpha -1)\right] -k\right\} z^{2}+\left\{ 4k-\left[ 4A+8\varepsilon
^{2}-(D+2l-2)^{2}+1\right] \right\} z+4\varepsilon ^{2}=0.$ Thus the
constant $k$ can be obtained as

\begin{equation}
k=A-\frac{(D+2l-2)^{2}-1}{4}\pm a\varepsilon ,\text{ \ }a=\sqrt{(1-2\alpha
)^{2}+(D+2l-2)^{2}-1}.
\end{equation}
In view of that, we can find four possible functions for $\pi (z)$ as
\begin{equation}
\pi (z)=-\frac{z}{2}\pm \left\{
\begin{array}{c}
\varepsilon -\left( \varepsilon -\frac{a}{2}\right) z,\text{ \ \ \ for \ \ }%
k=A-\frac{(D+2l-2)^{2}-1}{4}+a\varepsilon , \\
\varepsilon -\left( \varepsilon +\frac{a}{2}\right) z;\text{ \ \ \ for \ \ }%
k=A-\frac{(D+2l-2)^{2}-1}{4}-a\varepsilon .
\end{array}
\right.
\end{equation}
We must select

\begin{equation}
\text{\ }k=A-\frac{(D+2l-2)^{2}-1}{4}-a\varepsilon ,\text{ }\pi (z)=-\frac{z%
}{2}+\varepsilon -\left( \varepsilon +\frac{a}{2}\right) z,
\end{equation}
in order to obtain the polynomial, $\tau (z)=\widetilde{\tau }(z)+2\pi (z)$
having negative derivative as

\begin{equation}
\tau (z)=1+2\varepsilon -\left( 2+2\varepsilon +a\right) z,\text{ }\tau
^{\prime }(z)=-(2+2\varepsilon +a).
\end{equation}
We can also write the values of $\lambda =k+\pi ^{\prime }(z)$ and $\lambda
_{n}=-n\tau ^{\prime }(z)-\frac{n\left( n-1\right) }{2}\sigma ^{\prime
\prime }(z),$\ $n=0,1,2,...$ as

\begin{equation}
\lambda =A-\frac{(D+2l-2)^{2}-1}{4}-(1+a)\left[ \frac{1}{2}+\varepsilon %
\right] ,
\end{equation}
\begin{equation}
\lambda _{n}=n(1+n+a+2\varepsilon ),\text{ }n=0,1,2,...
\end{equation}
respectively. Additionally, using the definition of $\lambda =\lambda _{n}$
and solving the resulting equation for $\varepsilon ,$ allows one to obtain

\begin{equation}
\varepsilon =\frac{4(n+1)^{2}+(D+2l-2)^{2}-1+4(2n+1)\eta -4A}{8(n+1+\eta )},%
\text{ }\eta =\frac{-1+a}{2},
\end{equation}
from which we obtain the discrete energy levels

\begin{equation}
E_{nl}^{(D)}=-\frac{\hbar ^{2}}{32\mu b^{2}}\left[ \frac{%
4(n+1)^{2}+(D+2l-2)^{2}+4(2n+1)\eta -4A-1}{2(n+1+\eta )}\right] ^{2},\text{
\ }0\leq n,l<\infty
\end{equation}
where $n$ denotes the radial quantum number. It is found that $\Lambda $
remains invariant by mapping $\alpha \rightarrow 1-\alpha ,$ so do the bound
state energies $E_{nl}.$ An important quantity of interest for the
Manning-Rosen potential is the critical coupling constant $A_{c},$ which is
that value of $A$ for which the binding energy of the level in question
becomes zero. Using Eq. (26), in atomic units $\hbar ^{2}=\mu =Z=e=1,$

\begin{equation}
A_{c}=(n+1+\eta )^{2}-\eta (\eta +1)+\frac{(D+2l-2)^{2}}{4}-\frac{1}{4}.
\end{equation}

Let us now find the corresponding radial part of the wave function. Using $%
\sigma (z)$ and $\pi (z)$ in Eqs (57) and (61), we obtain

\begin{equation}
\phi (z)=z^{\varepsilon }(1-z)^{(\eta +1)/2},
\end{equation}
\begin{equation}
\rho (z)=z^{2\varepsilon }(1-z)^{2\eta +1},
\end{equation}

\begin{equation}
y_{nl}(z)=C_{n}z^{-2\varepsilon }(1-z)^{-(2\eta +1)}\frac{d^{n}}{dz^{n}}%
\left[ z^{n+2\varepsilon }(1-z)^{n+2\eta +1}\right] .
\end{equation}
The functions $\ y_{nl}(z)$ are, up to a numerical factor, are in the form
of\ Jacobi polynomials, i.e., $\ y_{nl}(z)\simeq P_{n}^{(2\varepsilon ,2\eta
+1)}(1-2z),$ valid physically in the interval $(0\leq r<\infty $ $%
\rightarrow $ $0\leq z\leq 1)$ [52]. Therefore, the radial part of the wave
functions can be found by substituting Eqs. (68) and (70) into $%
R_{nl}(z)=\phi (z)y_{nl}(z)$ as

\begin{equation}
g_{nl}(z)=N_{nl}z^{\varepsilon }(1-z)^{1+\eta }P_{n}^{(2\varepsilon ,2\eta
+1)}(1-2z),
\end{equation}
where $\varepsilon $ and $\Lambda $ are given in Eq. (65) and $N_{nl}$ is a
normalization constant. This equation satisfies the requirements; $%
R_{nl}(z)=0$ as $z=0$ $(r\rightarrow \infty )$ and $R_{nl}(z)=0$ as $z=1$ $%
(r=0).$ Therefore, the wave functions, $g_{nl}(z)$ in Eq. (71) is valid
physically in the closed interval $z\in \lbrack 0,1]$ or $r\in (0,\infty ).$
Further, the wave functions satisfy the normalization condition

\begin{equation}
\int\limits_{0}^{\infty }\left| g_{nl}(r)\right|
^{2}dr=1=b\int\limits_{0}^{1}z^{-1}\left| g_{nl}(z)\right| ^{2}dz,
\end{equation}
where $N_{nl}$ can be determined via

\begin{equation}
1=bN_{nl}^{2}\int\limits_{0}^{1}z^{2\varepsilon -1}(1-z)^{2\eta +2}\left[
P_{n}^{(2\varepsilon ,2\eta +1)}(1-2z)\right] ^{2}dz.
\end{equation}
Following Ref. ]36\}, we find the normalization constant
\begin{equation}
N_{nl}=\frac{1}{\sqrt{s(n)}},
\end{equation}
where
\[
s(n)=b(-1)^{n}\frac{\Gamma (n+2\eta +2)\Gamma (n+2\varepsilon +1)^{2}}{%
\Gamma (n+2\varepsilon +2\eta +2)}
\]
\begin{equation}
\times \sum\limits_{p,r=0}^{n}\frac{(-1)^{p+r}\Gamma (n+2\varepsilon
+r-p+1)(p+2\eta +2)}{p!r!(n-p)!(n-r)!\Gamma (n+2\varepsilon -p+1)\Gamma
(2\varepsilon +r+1)(n+2\varepsilon +r+2\eta +2)}.
\end{equation}
Therefore, we may express the normalized total wave functions as

\[
\psi ({\bf x})=N_{nl}r^{-(D-3)/2}e^{-\varepsilon r/b}(1-e^{-r/b})^{1+\eta
}P_{n}^{(2\varepsilon ,2\eta +1)}(1-2e^{-r/b})\exp (\pm il_{1}\theta _{1})
\]
\begin{equation}
\times N_{n_{D-1}}\sin (\theta _{D-1})^{l_{D-2}}P_{n_{D-1}}^{(\widetilde{%
\Lambda }_{D-2},\widetilde{\Lambda }_{D-2})}(\cos \theta
_{D-1}).\prod\limits_{j=2}^{D-2}N_{n_{j}}\left( \sin \theta _{j}\right)
^{l_{j-1}}P_{n_{j}}^{(\widetilde{\Lambda }_{j-1},\widetilde{\Lambda }%
_{j-1})}(\cos \theta _{j})
\end{equation}
To show the accuracy of our results, we calculate the energy eigenvalues for
various $n$ and $l$ quantum numbers with three different values of the
parameters $\alpha .$ As seen n Table 1, the energy eigenvalues for
different quantum numbers are obtained numerically for $D=2$ and $D=4$
cases.

>From Eq. (66), we have seen that two interdimensional states are degenerate
whenever [53]

\begin{equation}
(n,l,D)\rightarrow (n,l\pm 1,D\mp 2).
\end{equation}
Thus, a knowledge of $E_{nl}^{(D)}$ for $D=2$ and $D=3$ provides the
information necessary to find $E_{nl}^{(D)}$ for other higher dimensions.

For example, $E_{0,4}^{(2)}=E_{0,3}^{(4)}=E_{0,2}^{(6)}=E_{0,1}^{(8)}.$ This
is the same transformational invariance described for bound states of free
atoms and molecules [54,55] and demonstrates the existence of
interdimensional degeneracies among states of the confined Manning-Rosen
potential. \

As an example of incidental degeneracy, Table 1 presents the results of the
confined $2D$ and $4D$ Manning-Rosen energies at several radii of
confinement for various $n$ and $l$ states.

\section{Discussions}

In this work, we have utilized ${\rm NU}$ method to solve the $D$%
-dimensional ${\rm SE}$ for the Manning-Rosen model potential with the
angular momentum $l\neq 0$ states$.$ We have derived the binding energy
spectra in Eq. (66) and their corresponding wave functions in Eq. (71).

Let us study special cases. We have shown that for $\alpha =0$ $(1)$, the
present solution reduces to the one of the Hulth\'{e}n potential [16,18,19]:

\begin{equation}
V^{(H)}(r)=-V_{0}\frac{e^{-\delta r}}{1-e^{-\delta r}},\text{ }%
V_{0}=Ze^{2}\delta ,\text{ }\delta =b^{-1}
\end{equation}
where $Ze^{2}$ is the strength and $\delta $ is the screening parameter and $%
b$ is the range of potential. If the potential is used for atoms, the $Z$ is
identified with the atomic number. This can be achieved by setting $\eta =%
\frac{1}{2}(D+2l-3),$ hence, the energy for $l\neq 0$ states

\begin{equation}
E_{nl}=-\frac{\left[ 4A-(2n+D+2l-1)^{2}\right] ^{2}\hbar ^{2}}{32\mu
b^{2}(2n+D+2l-1)^{2}},\text{ \ }0\leq n,l<\infty .
\end{equation}
and for $s$-wave ($l=0)$ states
\begin{equation}
E_{n}=-\frac{\left[ A-(n+1)^{2}\right] ^{2}\hbar ^{2}}{8\mu b^{2}(n+1)^{2}},%
\text{ \ }0\leq n<\infty
\end{equation}
Essentially, these results coincide with those obtained by the Feynman
integral method [31] and the standard way [32,33], respectively.
Furthermore, if taking $b=1/\delta $ and identifying $\frac{A\hbar ^{2}}{%
2\mu b^{2}}$ as $Ze^{2}\delta ,$ we are able to obtain
\begin{equation}
E_{nl}=-\frac{\mu \left( Ze^{2}\right) ^{2}}{\hbar ^{2}}\left[ \frac{1}{%
(2n+D+2l-1)}-\frac{\hbar ^{2}\delta }{8Ze^{2}\mu }(2n+D+2l-1)\right] ^{2},
\end{equation}
which coincides with those of Refs. [16,18]. With natural units $\hbar
^{2}=\mu =Z=e=1,$ we have

\begin{equation}
E_{nl}=-\left[ \frac{1}{(2n+D+2l-1)}-\frac{(2n+D+2l-1)}{8}\delta \right]
^{2},
\end{equation}
which coincides with Refs. [16,33].

The corresponding radial wave functions are expressed as

\begin{equation}
R_{nl}(r)=N_{nl}e^{-\delta \varepsilon r}(1-e^{-\delta
r})^{(D+2l-1)/2}P_{n}^{(2\varepsilon ,D+2l-2)}(1-2e^{-\delta r})r^{-(D-3)/2},
\end{equation}
where

\begin{equation}
\varepsilon =\frac{2\mu Ze^{2}}{\hbar ^{2}\delta }\left[ \frac{1}{(2n+D+2l-1)%
}-\frac{\hbar ^{2}\delta }{8Ze^{2}\mu }(2n+D+2l-1)\right] ,\text{ }0\leq
n,l<\infty ,
\end{equation}
which coincides for the ground state with that given in Eq. (6) by
G\"{o}n\"{u}l {\it et al} [18]. In addition, for $\delta r\ll 1$ (i.e., $%
r/b\ll 1),$ the Hulth\'{e}n potential turns to become a Coulomb potential: $%
V(r)=-Ze^{2}/r$ with energy levels and wavefunctions:

\[
E_{nl}=-\frac{4\varepsilon _{0}}{(2n+D+2l-1)^{2}},\text{ }n=0,1,2,..
\]

\begin{equation}
.\varepsilon _{0}=\frac{Z^{2}\hbar ^{2}}{2\mu a_{0}^{2}},\text{ }a_{0}=\frac{%
\hbar ^{2}}{\mu e^{2}}
\end{equation}
where $\varepsilon _{0}=13.6$ $eV$ and $a_{0}$ is Bohr radius for the
Hydrogen atom. The wave functions are
\[
R_{nl}=N_{nl}\exp \left[ -\frac{8\mu Ze^{2}}{\hbar ^{2}}\frac{r}{\left(
2n+D+2l-1\right) }\right] r^{(D+2l-1)/2}P_{n}^{\left( \frac{2\mu Ze^{2}}{%
\hbar ^{2}\delta \left( 2n+D+2l-1\right) },D+2l-2\right) }(1+2\delta r)
\]
which coincide with Refs. [3,16,22].

\section{Cocluding Remarks}

In this work, we have extended the approximate solutions of the $l$-wave Schr%
\"{o}dinger equation with the Manning-Rosen potential to $D$-dimensions .
The special cases for $\alpha =0,1$ are discussed. The results are found to
be in good agreement with those obtained by other methods in $3D$ for short
potential range, small $\alpha $ and $l$ [36]$.$ These numerical solutions
have also extended to various dimensional space, $D=2$ and $D=4$ systems. We
have also studied two special cases for $l=0,$ $l\neq 0$ and Hulth\'{e}n
potential. The results we have ended up show that the NU method constitute a
reliable alternative way in solving the exponential potentials.

\acknowledgments
This research was partially supported by the Scientific and Technological
Research Council of Turkey.

\bigskip

\baselineskip= 2\baselineskip
\bigskip

\begin{table}[tbp]
\caption{Eigenvalues for $2p,3p,3d,4p,4d,4f,5p,5d,5f,5g,6p,6d,6f$ and $6g$
states in atomic units ($\hbar =\protect\mu =1)$ and for $\protect\alpha %
=0.75$ and $\protect\alpha =1.5,$ $A=2b.$}
\begin{tabular}{llllllll}
&  & $D=2$\tablenotetext[1]{Two-dimensional Schrodinger equation.}%
\tablenotemark[1] &  &  & $D=4$\tablenotetext[2]{Four-dimensional
Schrodinger equation.}\tablenotemark[2] &  &  \\
states & $1/b$ & $\alpha =0.75$ & $\alpha =0,1$ & $\alpha =1.5$ & $\alpha =$%
0.75 & $\alpha =$0,1 & $\alpha =1$.5 \\
\tableline$2p$ & $0.025$ & $-0.241087728$ & $-0.209898003$ & $-0.140949065$
& $-0.070734690$ & $-0.067988281$ & $-0.058898861$ \\
& $0.050$ & $-0.227946676$ & $-0.197925347$ & $-0.131737328$ & $-0.059344084$
& $-0.056953125$ & $-0.049054156$ \\
& $0.075$ & $-0.215173874$ & $-0.186304253$ & $-0.122836866$ & $-0.048952839$
& $-0.046894531$ & $-0.040109106$ \\
& $0.100$ & $-0.202769319$ & $-0.175034722$ & $-0.114247678$ & $-0.039560954$
& $-0.037812500$ & $-0.032063712$ \\
$3p$ & $0.025$ & $-0.074279113$ & $-0.067988281$ & $-0.051933432$ & $%
-0.030209821$ & $-0.029273358$ & $-0.026068346$ \\
& $0.050$ & $-0.062813564$ & $-0.056953125$ & $-0.042142549$ & $-0.020395577$
& $-0.019644452$ & $-0.017092049$ \\
& $0.075$ & $-0.052308602$ & $-0.046894531$ & $-0.033373420$ & $-0.012502916$
& $-0.011929608$ & $-0.010003237$ \\
& $0.100$ & $-0.042764227$ & $-0.037812500$ & $-0.025626042$ & $-0.006531840$
& $-0.006128827$ & $-0.004801908$ \\
$3d$ & $0.025$ & $-0.070734690$ & $-0.067988281$ & $-0.058898861$ & $%
-0.029833656$ & $-0.029273358$ & $-0.027228277$ \\
& $0.050$ & $-0.059344084$ & $-0.056953125$ & $-0.049054156$ & $-0.020047209$
& $-0.019644452$ & $-0.018176769$ \\
& $0.075$ & $-0.048952839$ & $-0.046894531$ & $-0.040109106$ & $-0.012199670$
& $-0.011929608$ & $-0.010947973$ \\
$4p$ & $0.025$ & $-0.031448122$ & $-0.029273358$ & $-0.023381941$ & $%
-0.014180352$ & $-0.013773389$ & $-0.012357598$ \\
& $0.050$ & $-0.021545731$ & $-0.019644452$ & $-0.014606136$ & $-0.006296995$
& $-0.006019483$ & $-0.005072360$ \\
& $0.075$ & $-0.013510134$ & $-0.011929608$ & $-0.007885467$ & $-0.001570215$
& $-0.001429639$ & $-0.000978205$ \\
$4d$ & $0.025$ & $-0.030209821$ & $-0.029273358$ & $-0.026068346$ & $%
-0.014011823$ & $-0.013773389$ & $-0.012892982$ \\
& $0.050$ & $-0.020395577$ & $-0.019644452$ & $-0.017092049$ & $-0.006162813$
& $-0.006019483$ & $-0.005494347$ \\
& $0.075$ & $-0.012502916$ & $-0.011929608$ & $-0.010003237$ & $-0.001492711$
& $-0.001429639$ & $-0.001204122$ \\
$4f$ & $0.025$ & $-0.029833656$ & $-0.029273358$ & $-0.027228277$ & $%
-0.013929374$ & $-0.013773389$ & $-0.013182139$ \\
& $0.050$ & $-0.020047209$ & $-0.019644452$ & $-0.018176769$ & $-0.006097355$
& $-0.006019483$ & $-0.005724889$ \\
& $0.075$ & $-0.012199670$ & $-0.011929608$ & $-0.010947973$ & $-0.001455297$
& $-0.001429639$ & $-0.001333163$ \\
$5p$ & $0.025$ & $-0.014732070$ & $-0.013773389$ & $-0.011100961$ & $%
-0.007127957$ & $-0.006916484$ & $-0.006175251$ \\
$5d$ & $0.025$ & $-0.014180352$ & $-0.013773389$ & $-0.012357598$ & $%
-0.006506751$ & $-0.006392207$ & $-0.005967020$ \\
$5f$ & $0.025$ & $-0.014011823$ & $-0.013773389$ & $-0.012892982$ & $%
-0.006465489$ & $-0.006392207$ & $-0.006113207$ \\
$5g$ & $0.025$ & $-0.013929374$ & $-0.013773389$ & $-0.013182139$ & $%
-0.006440958$ & $-0.006392207$ & $-0.006204004$ \\
$6p$ & $0.025$ & $-0.006866319$ & $-0.006392207$ & $-0.005056211$ & $%
-0.002734814$ & $-0.002635101$ & $-0.002286461$ \\
$6d$ & $0.025$ & $-0.005435481$ & $-0.006392207$ & $-0.005695750$ & $%
-0.002691847$ & $-0.002635101$ & $-0.002424502$ \\
$6f$ & $0.025$ & $-0.006506751$ & $-0.006392207$ & $-0.005967020$ & $%
-0.002670817$ & $-0.002635101$ & $-0.002499036$ \\
$6g$ & $0.025$ & $-0.006465489$ & $-0.006392207$ & $-0.006113207$ & $%
-0.002658317$ & $-0.002635101$ & $-0.002545374$%
\end{tabular}
\end{table}


\newpage

\begin{references}
\bibitem{}  L.I. Schiff, Quantum Mechanics 3rd edn. (McGraw-Hill Book Co.,
New York, 1968).

\bibitem{}  L.D. Landau and E.M. Lifshitz, Quantum Mechanics,
Non-relativistic Theory, 3rd edn. (Pergamon, New York, 1977).

\bibitem{}  M.M. Neito, Am. J. Phys. 47 (1979) 1067.

\bibitem{}  \c{S}. Erko\c{c} and R. Sever, Phys. Rev. D 30 (1984) 2117; D 33
(1986) 588; Phys. Rev. A 37 (1988) 2687.

\bibitem{}  M.L. Sage, Chem. Phys. 87 (1984) 431; M. Sage and J. Goodisman,
Am. J. Phys. 53 (1985) 350.

\bibitem{}  S.-H. Dong, Appl. Math. Lett. 16 (2003) 199.

\bibitem{}  S. Ikhdair and R. Sever, J. Mol. Struct.-Theochem 806 (2007) 155.

\bibitem{}  S.M. Ikhdair and R. Sever, preprint quant-ph/0611065, to appear
in J. Mol. Struct.-Theochem (2008).

\bibitem{}  C.L. Pekeris, Phys. Rev. 45 (1934) 98.

\bibitem{}  C. Berkdemir, Nucl. Phys. A 770 (2006) 32.

\bibitem{}  W.-C. Qiang and S.-H. Dong, Phys. Lett. A 363 (2007) 169.

\bibitem{}  C. Berkdemir and J. Han, Chem. Phys. Lett. 409 (2005) 203.

\bibitem{}  W.-C. Qiang and S.-H. Dong, Phys. Lett. A 363 (2007) 169.

\bibitem{}  C. Berkdemir, A. Berkdemir and J. Han, Chem. Phys. Lett. 417
(2006) 326.

\bibitem{}  P.M. Morse, Phys. Rev. 34 (1929) 57.

\bibitem{}  S.M. Ikhdair and R. Sever, J. Math. Chem. 42 (3) (2007) 461.

\bibitem{}  M. \c{S}im\c{s}ek and H. E\u{g}rifes, J. Phys. A: Math. Gen. 37
(2004) 4379.

\bibitem{}  B. G\"{o}n\"{u}l and \.{I}. Zorba, Phys. Lett. A 269 (2000) 83;
R.L. Greene and C. Aldrich, Phys. Rev. A 14 (1976) 2363.

\bibitem{}  L. Hulth\'{e}n, Ark. Mat. Astron. Fys. A 28 (1942) 5.

\bibitem{}  \"{O}. Ye\c{s}ilta\c{s}, Phys. Scr. 75 (2007) 41.

\bibitem{}  S.M. Ikhdair and R. Sever, Ann. Phys. (Berlin) 16 (2007) 218.

\bibitem{}  S.M. Ikhdair and R. Sever, Int. J. Theor. Phys. 46 (2007) 1643;
46 (2007) 2384.

\bibitem{}  S.M. Ikhdair and R. Sever, preprint quant-ph/0605045], to appear
in the Int. J. Mod. Phys. E.

\bibitem{}  C. Berkdemir, Am. J. Phys. 75 (2007) 81.

\bibitem{}  Y.-F. Cheng and T.-Q. Dai, Phys. Scr. 75 (2007) 274; Chinese J.
Phys. 45 (5) (2007) 480.

\bibitem{}  S.M. Ikhdair and R. Sever, Int. J. Mod. Phys. C 18 (10) (2007)
1571; Centr. Eur. J. Phys. 5 (4) (2007) 516; preprint quant-ph/0702052 to
appear in the Centr. Eur. J. Phys.; S.M. Ikhdair, preprint quant-ph/0703042,
to appear in the Chinese J. Phys.

\bibitem{}  W.-C. Qiang, Chin. Phys. 12 (2003) 1054; 13 (2004) 575.

\bibitem{}  L.-Z. Yi, Y.-F. Diao, J.-Y. Liu and C.-S. Jia, Phys. Lett. A 333
(2004) 212.

\bibitem{}  N. Rosen and P.M. Morse, Phys. Rev. 42 (1932) 210.

\bibitem{}  M.F. Manning, Phys. Rev. 44 (1933) 951.

\bibitem{}  A. Diaf, A. Chouchaoui and R.L. Lombard, Ann. Phys. (Paris) 317
(2005) 354.

\bibitem{}  S.-H. Dong and J. Garcia-Ravelo, Phys. Scr. 75 (2007) 307.

\bibitem{}  W.-C. Qiang and S. H. Dong, Phys. Lett. A 368 (2007) 13.

\bibitem{}  C.-S. Jia {\it et al}., J. Phys. A: Math. Gen. 37 (2004) 11275;
\ C.-S. Jia {\it et al}., Phys. Lett. A 311 (2003) 115.

\bibitem{}  H. E\u{g}rifes, D. Demirhan and F. B\"{u}y\"{u}kk\i l\i \c{c},
Phys. Lett. A 275 (2000) 229.

\bibitem{}  S.M. Ikhdair, arXiv:0801.1500, submitted to Int. J. Mod. Phys. E.

\bibitem{}  A.F. Nikiforov and V.B. Uvarov, Special Functions of
Mathematical Physics (Birkhauser, Bassel, 1988).

\bibitem{}  R.J. Le Roy and R.B. Bernstein, J. Chem. Phys. 52 (1970) 3869.

\bibitem{}  J. Cai, P. Cai and A. Inomata, Phys. Rev. A 34 (1986) 4621.

\bibitem{}  S.M. Ikhdair and R. Sever, arXiv: 0704.0573, to appear in the
Cent. E. J. Phys.

\bibitem{}  S.M. Ikhdair, preprint quant-ph/0703042, to appear in the
Chinese J. Phys.; preprint quant-ph/0703008, to appear in the Int. J. Mod.
Phys. C.

\bibitem{}  S.M. Ikhdair and R. Sever, quant-ph/0703131 to appear in the
Cent. E. J. Phys.

\bibitem{}  J.D. Louck and W.H. Shaffer, J. Mol. Spec. 4 (1960) 285; J.D.
Louck, 4 (1960) 298; 4 (1960) 334.

\bibitem{}  J.D. Louck, Theory of Angular Momentum in D-Dimensional Space,
Los Alamos Scientific Laboratory monograph LA-2451 (LASL, Los Alamos, 1960).

\bibitem{}  J.D. Louck and H.W. Galbraith, Rev. Mod. Phys. 48 (1976) 69.

\bibitem{}  A. Chatterjee, Phys. Rep. 186 (1990) 249.

\bibitem{}  A. Erd\'{e}lyi, Higher Transcendental Functions, Vol. 2 (McGraw
Hill, New York, 1953).

\bibitem{}  B.H. Bransden and C.J. Joachain, Quantum Mechanics, 2nd Ed.
(Pearson Education Limited, Great Britain, 2000) pp. 336.

\bibitem{}  L.-Y. Wang, X.-Y. Gu, Z.-Q. Ma and S.-H. Dong, Found. Phys.
Lett. 15 (2002) 569; S.-H. Dong, App. Math. Lett. 16 (2003) 199.

\bibitem{}  S.H. Dong, Phys. Scr. 64 (2001) 273; 65 (2002) 289.

\bibitem{}  F. Yasuk, A. Durmus and I. Boztosun, J. Math. Phys. 47 (2006)
082302; A. Durmus and F. Yasuk, J. Chem. Phys. 126 (2007) 074108.

\bibitem{}  M. Abramowitz and I.A. Stegun, Handbook of Mathematical
Functions (Dover, New York, 1964).

\bibitem{}  H.E. Montgomery, JR, N. A. Aquino and K. D. Sen, Int. J. Quantum
Chem. 107 (2007) 798.

\bibitem{}  D.R. Herrick, J. Math. Phys. 16 (1975) 281; D. R. Herrick and F.
H. Stillinger, Phys. Rev. 11 (1975) 42.

\bibitem{}  D.D. Fratz and D.R. Herschbach, J. Chem. Phys. 92 (1990) 6668.
\end{references}
\end{document}